\documentclass[%
 reprint,
 amsmath,amssymb,
prb,
floatfix,
]{revtex4-2}
\usepackage{empheq}
\usepackage{bbm}
\usepackage{float}
\usepackage[normalem]{ulem}
\usepackage{lipsum}
\usepackage{makecell}
\usepackage{mathtools, cuted, relsize}
\usepackage{esvect}
\usepackage{soul, xcolor}
\usepackage[caption = false]{subfig}
\usepackage{graphicx}
\usepackage{dcolumn,multirow, tabularx}
\usepackage{bm}
\usepackage{hyperref}
\usepackage{physics}
\usepackage{slashed}
\usepackage{balance}
\usepackage[mathlines]{lineno}
\RequirePackage{fix-cm}

\newcommand{\size}[2]{{\fontsize{#1}{0}\selectfont#2}}

\begin{document}

\preprint{APS/123-QED}

\title{Field Theoretical Study of Disorder in Non-Hermitian Topological Models}
\author{Anouar Moustaj}
\affiliation{
Institute for Theoretical Physics, Utrecht University, Princetonplein 5, 3584CC Utrecht, The Netherlands}%
\author{Lumen Eek}
 \affiliation{
Institute for Theoretical Physics, Utrecht University, Princetonplein 5, 3584CC Utrecht, The Netherlands}%
\author{Cristiane Morais Smith}
\affiliation{
Institute for Theoretical Physics, Utrecht University, Princetonplein 5, 3584CC Utrecht, The Netherlands}%

\date{April 25, 2022}

\begin{abstract}
Non-Hermitian systems have provided a rich platform to study unconventional topological phases. These phases are usually robust against external perturbations that respect certain symmetries of the system. In this work, we provide a new method to analytically study the effect of disorder, using tools from quantum field theory applied to discrete models around phase-transition points. We investigate two different one-dimensional models, the paradigmatic non-Hermitian SSH model and a $s$-wave superconductor with imbalanced pairing. These analytic results are compared to numerical simulations in the discrete models. It is found that the systems are driven from a topological to a trivial phase in the same way.

\end{abstract}

\maketitle


\paragraph*{Introduction.}
Systems described by non-Hermitian Hamiltonians have attracted great interest in the last few years. Usually, any observable has to be represented by a Hermitian operator, such that its eigenvalues are real. However, when one considers more complicated systems, an effective description using non-Hermitian Hamiltonians might be useful.
A typical example arises when studying transport phenomena \cite{Datta, FARID}. In this case, the effective description of an open system results in a non-Hermitian Hamiltonian, giving rise to states with finite lifetimes and complex energies. More generally, while an isolated system provides an ideal platform to understand its main characteristics, a more realistic description should include coupling to its environment.  Many realizations of non-Hermitian models find place in mechanical, atomic, and optical systems, in which gain and loss can be controlled \cite{Bandreseaar4005, Cerjan, Helbig, Weiman2017}. This provides a solid experimental ground to theoretical studies, on which the predictions can be tested.

The recent spark of interest for non-Hermitian Hamiltonians arose when it was realized that a class of models allowed for an extension of the topological classification based on protecting symmetries, which exists in the Hermitian case \cite{AltZNB,Gong2018,edgestatesnh,toponumber2D,geomeaninghalfint, NHchernbands, PeriodicTableNHSYM}. There exist several differences between the Hermitian and non-Hermitian classifications. For example, the bulk-boundary correspondence usually breaks down for non-Hermitian topological phases, and has to be replaced by a more appropriate measure of boundary phenomena, such as biorthogonal polarization \cite{Kunst2018}, or through a non-Bloch bulk-boundary correspondence \cite{Kunst,LonghiNBPTNHSE,AnatomyLee}. Furthermore, many models exhibit the non-Hermitian skin-effect \cite{NHBMTP,Okuma2020}. In addition, a recent study of the critical behaviour of topological phase transitions in non-Hermitian models has revealed unconventional scaling exponents, suggesting that these systems lie in different universality classes than their Hermitian counterparts \cite{Rodrigo}.\\
\indent In this Letter, we provide a method to study the effect of disorder on the strictly non-Hermitian topological characterisations of these systems. Our framework is valid for systems that exhibit a band closing that changes the topology from a line-gap to a point-gap. We develop a systematic method to implement disorder by extending the discrete system to its continuous description around the band-closing point characterising the topological phase transitions. We start our study by considering a general, discrete two-band model around the critical point, and write its continuous version. This provides us with a (1+1) dimensional field theory, which only behaves properly after selecting a single frequency component of the temporal part, as proposed by Kawabata et al.~\cite{TFTKawabata2021}. The effective (1+0) dimensional theory turns out to be topological when coupled to a gauge field. We then introduce disorder and investigate it perturbatively. The procedure is applied to the non-Hermitian SSH model and a non-Hermitian $s$-wave superconductor. The results are compared with numerical calculations of the discrete models. We find that strong disorder drives the system from a topological to a trivial phase in the same way. 
\paragraph*{Topological Field Theory of Two-Band Models.}
We consider Hamiltonians of the form
\begin{equation}\label{eq:genH}
    \hat{H}(k)=\sum_k \hat{\mathbf{c}}^\dagger_k\big[\mathbf{f}(k)\cdot\boldsymbol{\sigma}\big]\hat{\mathbf{c}}_k,
\end{equation}
where $\boldsymbol{\sigma}=(\sigma_1,\sigma_2,\sigma_3)$ is a vector of Pauli matrices, and the complex-valued function $\mathbf{f}(k)$ depends on the microscopic properties of the system. The annihilation operators $\mathbf{c}_k=(c_{k,A},c_{k,B})$ are bipartite, and reflect the two-band structure of the system. An important assumption about the Hamiltonian in Eq.~\eqref{eq:genH} is that it has at least one line-gap closing that changes the gap topology from a line-gap to a point-gap (see the supplemental material (SM) for an example of such gap closings \cite{SM}). The resulting continuum Hamiltonian is given in momentum space by
\begin{equation}
    \hat{H}(k) = \int \frac{\dd k}{2\pi} \hat{\boldsymbol{\Psi}}^\dagger(k) \left[ (\boldsymbol{\alpha}\cdot\boldsymbol{\sigma})k + \boldsymbol{\beta}\cdot\boldsymbol{\sigma} \right] \hat{\boldsymbol{\Psi}}(k),\label{eq:conH}
\end{equation}
where $\boldsymbol{\alpha}$ and $\boldsymbol{\beta}$ result from the linear expansion of $\mathbf{f}(k)$ around the line-gap closing and the field operator $\boldsymbol{\Psi}(x) = (\psi_A(x),\psi_B(x))^T$ obeys anticommutation relations $\{\psi_A(x_i),\psi_B(x_j)\}=\delta_{AB}\delta(x_i-x_j)$. The action associated to this Hamiltonian is given by
\begin{equation}
    S = \int \dd t \dd x \boldsymbol{\Psi}^\dagger(x) \left[ i \partial_t + i (\boldsymbol{\alpha}\cdot\boldsymbol{\sigma}) \partial_x - \boldsymbol{\beta}\cdot\boldsymbol{\sigma} \right] \boldsymbol{\Psi}(x), \notag
\end{equation}
where we have set $\hbar = 1$.

Following Kawabata et al.~\cite{TFTKawabata2021}, we consider a field theory in (1+1) dimensions and discard the temporal degree of freedom, which makes the theory ill-defined. The corresponding action is
\begin{align}
    S^{(E)} &= \int \dd x \boldsymbol{\Psi}_E^\dagger(x) \left[ E + i (\boldsymbol{\alpha}\cdot\boldsymbol{\sigma}) \partial_x - \boldsymbol{\beta}\cdot\boldsymbol{\sigma} \right] \boldsymbol{\Psi}_E(x), \notag 
\end{align}
where the index $E$ denotes a fixed energy. The non-Hermitian topological character of the system naturally arises by coupling the matter fields to a background $U(1)$ gauge field $A$ by virtue of minimal substitution: $\partial_x \to \partial_x - i A$,
\begin{equation}
    S^{(E)}[A] = \int \dd x \boldsymbol{\Psi}_E^\dagger(x) \left[ G^{-1}_{0,E} + (\boldsymbol{\alpha}\cdot\boldsymbol{\sigma})A \right] \boldsymbol{\Psi}_E(x), \notag
\end{equation}
where we introduced the (inverse) bare Green function $G^{-1}_{0,E}$. The vacuum-to-vacuum transition amplitude is then given by
\begin{align}
    Z_E[A] = \int \hspace{-2pt} \mathcal{D}\boldsymbol{\Psi}_E^\dagger\mathcal{D}\boldsymbol{\Psi}_E e^{iS^{(E)}} \notag  
    = \text{Det} \hspace{-2pt} \left[ -i G^{-1}_{0,E} - i(\boldsymbol{\alpha}\cdot\boldsymbol{\sigma})A\right]. \notag
\end{align}
Here, $\text{Det}[...]$ denotes a determinant over both coordinate and spinor spaces, while $\text{det}[...]$ only accounts for spinor space. The same convention holds for taking traces.
We now shift our attention to the effective action, defined through $\exp{i S_\text{eff}} = Z_E[A]/Z_E[0]$. The gauge field $A$ is assumed to be small in magnitude, such that we can probe the system in linear response. At this order, the effective action is given by
\begin{align}
    S_\text{eff} &= -i \Tr  \left[  G_{0,E}  (\boldsymbol{\alpha}\cdot\boldsymbol{\sigma})A \right] \notag \\
    &= \int \frac{\dd k}{2 \pi i} \tr \left[ G_{0,E}(k)  (\boldsymbol{\alpha}\cdot\boldsymbol{\sigma}) \right] \int \dd x A(x) \notag \\
    &\equiv \mathcal{W}(E) \int \dd x A(x), \label{eq:effaction}
\end{align}
where in the second line the trace over coordinate space was explicitly taken and in the third line we introduced the `energy vorticity' $\mathcal{W}(E)$ (more details are given in the SM \cite{SM}). The energy vorticity captures the response of the system to the applied gauge field. It turns out that it is exactly equal to the winding of the complex energy spectrum around the point $E$, making it a topological invariant. Furthermore, $\mathcal{W}(E)$ appears as the current that results from the coupling to the gauge field~\cite{TFTKawabata2021}. This generically gives a proper physical interpretation of this purely non-Hermitian winding number, and in the case of open boundary conditions, is an indicator of the appearance of the non-Hermitian skin-effect \cite{LonghiNBPTNHSE, NHBMTP, Okuma2020, geomeaninghalfint, Gong2018, LonghiDeform, HermnonHermInv}.

\paragraph*{Effect of Disorder.}
We now introduce an averaged disorder to the model to see how it affects the non-Hermitian topological phases. From Eq.~\eqref{eq:effaction}, we see that this means that a modification will take place in the bare Green's function through the introduction of disorder in the single particle Hamiltonian $H = H_0 + V$, where $H_0$ is an unperturbed Hamiltonian, and $V$ incorporates disorder into the system. Here, we will consider the disorder to be a deviation from a zero-average configuration. We replace $V(x)$ with $\delta V(x) = V(x) - \overline{V(x)}$, with $V(x)$ a disorder potential, assumed to be of the form $V(x) = \sum_{i=1}^{N_{imp}} U(x-x_i)$,
where $N_{imp}$ is the number of impurities and $U(x)$ is an arbitrary function capturing the nature of the disorder. Thus, the spatial averaged disorder $\overline{V(x)}$ is defined as an average procedure over all $x_j$. The (spatial) action in presence of disorder then reads
\begin{equation}
    S^{(E)} = \int \dd x \boldsymbol{\Psi}_{E}^\dagger(x) \left[ E - H_0 - \delta V \right] \boldsymbol{\Psi}_{E}(x).\notag
\end{equation}
The Green function can be expressed as a functional integral over fermionic fields
\begin{equation}
    G_E(x,y) = -\frac{1}{Z_E} \int \mathcal{D}\boldsymbol{\Psi}_E^\dagger\mathcal{D}\boldsymbol{\Psi}_E \boldsymbol{\Psi}_E(x) \boldsymbol{\Psi}_E^\dagger(y) e^{iS^{(E)}}. \label{eq:G}
\end{equation}
The disorder contribution in the exponential of Eq.~\eqref{eq:G} is expanded, and then a disorder average is taken \cite{Coleman}. This procedure eliminates all terms odd in $\overline{\delta V(x)}$, and the result is the Dyson equation for the disorder averaged Green function:
\begin{align}
    \overline{G_E(x,y)} &= G_{0,E}(x,y) \\ &+ \int dx'dx''G_{0,E}(x,x')\Sigma_E(x',x'')\overline{G_{E}(x'',y)},\notag
\end{align}
where the ``self-energy'', at a Born-approximation level, is given by 
\begin{equation}
    \Sigma_E(x,y)=-G_{0,E}(x,y)\overline{\delta V(x)\delta V(y)}. \label{eq:selfenergy}
\end{equation}
For this approximation to be valid, we require the disorder potential to be weak with respect to the eigenvalues of the unperturbed system. This will then be used as the starting point in the iterative calculation of the self-consistent Born approximation (SCBA), see SM. Note that disorder averaging reinstates translational invariance, $\Sigma_E(x,y) = \Sigma_E(x-y)$. Solving the Dyson equation in momentum space yields the modified Green function
\begin{equation}    
    \overline{G_E(k)} = [1-\Sigma_E(k)]^{-1} G_{0,E}(k). \label{eq:Gk}
\end{equation}
In the presence of weak disorder Eq.~\eqref{eq:effaction} and Eq.~\eqref{eq:Gk} can be combined such that the disorder averaged winding number takes the form
\begin{equation}
    \overline{\mathcal{W}(E)}=\int\frac{\dd k}{2\pi i}\tr\left[(1-\Sigma_E(k))^{-1} G_{0,E}(k)(\boldsymbol{\alpha}\cdot\boldsymbol{\sigma})\right].
\end{equation}
We consider a delta-function disorder $U(x-x_j) = U_0 \delta(x-x_j)$, where $U_0$ represents the disorder strength (scaled with units of length). This results in $\overline{\delta V(x) \delta V(y)} \approx U_0^2 n_i \delta(x-y)$, where the impurity density $n_i = N_{imp}/L$ was introduced, with $L$ the size of the system (see SM \cite{SM}). Combining this with Eq.~\eqref{eq:selfenergy} yields the self energy for the delta-function disorder averaged system in momentum space
\begin{equation}
    \Sigma_E(k) = - U_0^2 n_i \int \frac{\dd q}{2\pi} G_0(q).\label{eq:selfenergydelta}
\end{equation}
From Eq.~\eqref{eq:selfenergydelta}, we observe that the self-energy is momentum independent, which makes it possible to evaluate the energy vorticity analytically.

\begin{figure}
    \centering
    \includegraphics[width=0.9\columnwidth]{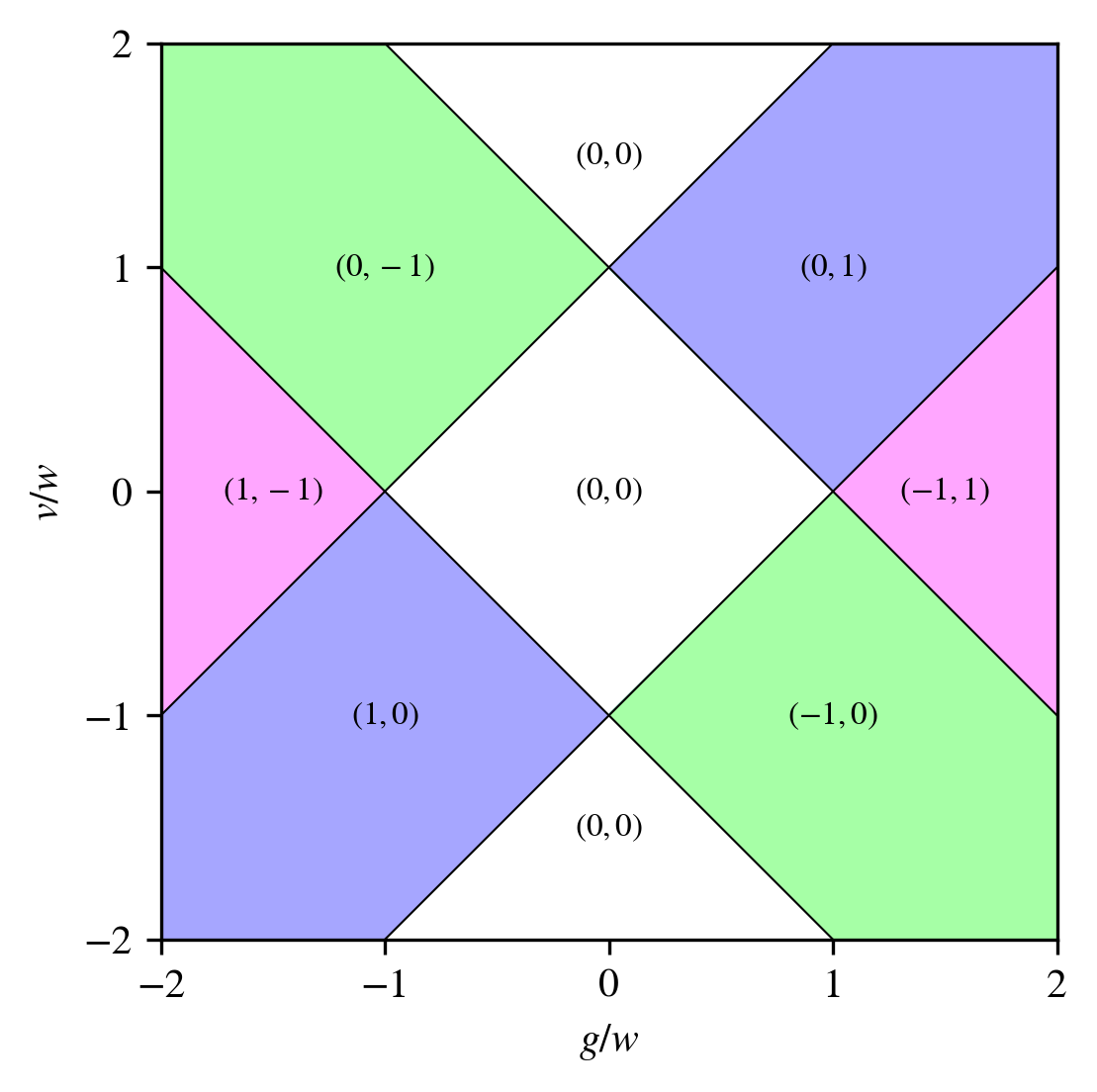}
    \caption{Phase diagram for the non-Hermitian SSH model, obtained from the non-Hermitian winding numbers $(\mathcal{W}_+(0),\mathcal{W}_-(0))$. The invariants come in pairs, where the first one is calculated around the gap closing point $k=0$ and the second one around the point $k=\pi/a$.}
    \label{fig:phaseSSH}
\end{figure}

\paragraph*{Non-Hermitian SSH model.} A paradigmatic model to study non-Hermitian topological matter in one dimension is the non-Hermitian SSH model for fermions. The SSH model describes a bipartite one-dimensional chain with A and B sites obeying a sub-lattice symmetry (see Fig.~S1 of the SM \cite{SM}). Here, we consider the non-Hermitian SSH model with non-reciprocal intracell hopping. The corresponding Hamiltonian reads
\begin{align}
    \hat{H} &= (v-g) \sum_{j=1}^N c^\dagger_{A,j} c_{B,j} + (v+g) \sum_{j=1}^N c^\dagger_{B,j} c_{A,j} \notag \\
    &+ w \sum_{j=1}^N \left( c^\dagger_{B,j} c_{A,j+1} + c^\dagger_{A,j+1} c_{B,j}\right), \notag
\end{align}
where the intra- and inter-cell hopping are denoted by $v$ and $w$, respectively. Moreover, the parameter $g$ introduces non-reciprocity in the intracell hopping. $N$ denotes the number of unit cells. 
The dispersion relation follows from diagonalising the Hamiltonian in momentum space, yielding
\begin{equation}
    E_\pm(k) = \pm \sqrt{w^2 + v^2 - g^2 + 2wv \cos k -2iwg \sin k}. \notag
\end{equation}
It supports line-gap closings at momenta $k = 0,\pi$. Expanding the Hamiltonian around these points then gives $\boldsymbol{\alpha} = (0,\pm w,0)$ and $\boldsymbol{\beta} = (v\pm w,-ig,0)$, where the $\pm$ denotes the expansions around $k=0$ and $k=\pi$, respectively. Evaluating the energy vorticity for the non-Hermitian SSH model yields
\begin{align}
    \mathcal{W}_\pm (E) = \mp \frac{1}{2} \big[ \text{sgn} \left( \gamma - \eta\right)  + \text{sgn} \left( \gamma + \eta\right) \big], \label{eq:SSHWinding}
\end{align}
with $\gamma = g/w$, $\eta=\Re \sqrt{M_\pm^2-\mathcal{E}^2}$, $M_\pm = (v\pm w)/w$ and $\mathcal{E}  = E/w$. The full model is described by the combination of these two Dirac models. The components $\mathcal{W}_\pm(E)$ relate to the full invariant as $\mathcal{W}(E) = \mathcal{W}_+(E) + \mathcal{W}_-(E)$. We remark that the vorticity is equal to the difference of the two half-integer windings around the exceptional points \cite{geomeaninghalfint}, $\mathcal{W}(0) = \nu_1 - \nu_2$.
These windings are defined through
\begin{align}
    \nu_{j} = \int \frac{\dd k}{2\pi} \frac{d}{dk} \arctan \left[\frac{\Re f_2 - (-1)^j \Im f_1}{\Re f_1 + (-1)^j \Im f_2}\right], \notag 
\end{align}
where the functions $f_{1,2}$ are introduced in Eq.~\eqref{eq:genH}. Note that $\mathcal{W}(E)$ is evaluated at $E=0$ because the non-Hermitian SSH model possesses sublattice symmetry. Fig.~\ref{fig:phaseSSH} shows the phase diagram of the non-Hermitian SSH model. This invariant renders all phases that are adiabatically connected to the Hermitian model ($g=0$) indistinguishable from each other, which is a result of the purely non-Hermitian nature of the energy vorticity $\mathcal{W}_\pm(E)$.

\begin{figure*}[!hbt]
    \centering
    \includegraphics[width=\textwidth]{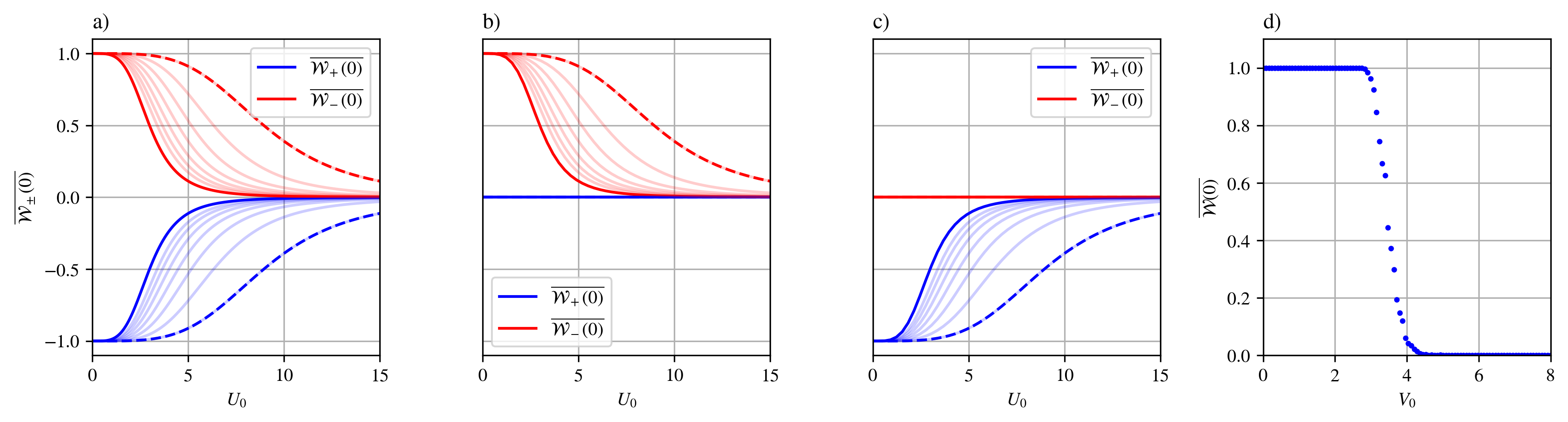}
    \caption{Disorder-driven topological to trivial phase transition in the non-Hermitian SSH model. (a-c) Average energy vorticity $\mathcal{W}_+(0)$ and $\mathcal{W}_-(0)$ as a function of $U_0$ starting from the clean phase, with parameters (a) $(v,w,g,n_i)=(0.5,1,2,0.025)$, (b) $(v,w,g,n_i)=(1,1,1,0.025)$ and (c) $(v,w,g,n_i)=(-1,1,1,0.025)$, corresponding to three non-Hermitian topologically distinct phases. A transition from the topological to the trivial phase is clearly observed. The dashed line corresponds to the Born approximation calculation, while the consecutive lines correspond to the multiple iterations of the SCBA. Note that the energy vorticity shows non-integer values around the critical point because it is an averaged quantity. (d) Numerically calculated average energy vorticity for the discrete non-Hermitian SSH model upon inclusion of on-site disorder. The calculations are done for $N=100$ cells, $(v,w,g) = (1,1,1)$ and averages were taken over $100$ disorder realisations.}
    \label{fig:disorderwide}
\end{figure*}

Upon including disorder in the non-Hermitian SSH model, the self-energy, defined through Eq.~\ref{eq:selfenergydelta}, reads
\begin{equation}
    \Sigma_E(k) =  -\frac{U^2_0 n_i}{2w}\begin{pmatrix}
    0 & -\text{sgn}(\gamma+M_\pm) \\ \text{sgn}(\gamma-M_\pm) & 0
    \end{pmatrix}, \notag
\end{equation}
from which the corrected Green function readily follows through Eq.~\eqref{eq:Gk}. Evaluating Eq.~\eqref{eq:effaction} using the corrected Green function then yields the energy vorticity for the disorder averaged non-Hermitian SSH model within the Born approximation (see SM for a full derivation \cite{SM}),
\begin{equation}
    \overline{\mathcal{W}_\pm(0)} =  \frac{\mathcal{W}_\pm(0)}{1+\left(\frac{U_0^2 n_i}{2w}\right)^2 \text{sgn}(\gamma-M_\pm) \text{sgn}(\gamma+M_\pm)}.\label{eq:avwind}
\end{equation}
Starting from this expression, we can initiate the calculation of the SCBA. Several iterations of the SCBA, corresponding to different regions in Fig.~\ref{fig:phaseSSH}, are plotted in Fig.~\ref{fig:disorderwide}, indicating that the curve converges to a sharper transition as the number of iterations grows. We observe that $\mathcal{W}_+(0)$ ($\mathcal{W}_-(0)$) is driven from minus (plus) one to zero [Fig.~\ref{fig:disorderwide}(a)]. For phases where either $\mathcal{W}_\pm(0)$ is already zero in the unperturbed system, they remain zero [Fig.~\ref{fig:disorderwide}(b,c)], as can be inferred from Eq.~\eqref{eq:avwind}. The disorder implementations in the continuum and discrete models are different in nature. It is therefore sensible to make a comparison between the models based on the spatial two-point correlations of the disorder potential. In the continuous model, it takes the form $\overline{\delta V(x) \delta V(y)}=U_0^2n_i\delta(x-y)$, while in the discrete model, it is $\overline{V_iV_j}=(V_0^2/3)\delta_{ij}$. The impurity density $n_i$ influences the location and sharpness of the transition. We use $n_i=0.025$ to compare the transitions in Fig.~\ref{fig:disorderwide}, and show how the energy vorticity depends on the impurity density $n_i$ in the SM \cite{SM}. Note that the impurity
strengths have different dimensions in both models.
\paragraph*{Non-Hermitian $s$-wave Superconducting Chain.}
Now, consider a spinful ($S=1/2)$ superconducting chain given by the following Hamiltonian:
\begin{align}
    H &=  w\sum_{j,s}(c_{j,s}^\dagger c_{j+1,s}+\text{h.c.})-\mu\sum_{j,s}(c_{js}^\dagger c_{js}-1) \notag\\
    &+\Delta\sum_{j}(c_{j,\uparrow}^\dagger c_{j+1,\downarrow}^\dagger+\text{h.c.}) + g\sum_{j}(c_{j,\uparrow}^\dagger c_{j,\downarrow}^\dagger-c_{j,\downarrow} c_{j,\uparrow}), \notag
\end{align}
where $s=\uparrow,\downarrow$ represents spin, $w$ is the hopping parameter, $\mu$ the chemical potential, $\Delta$ the nearest-neighbour superconducting pairing strength, and the non-Hermiticity is introduced through the parameter $g$, representing an imbalance in the on-site superconducting pairing. The Hermitian model exhibits nontrivial topological phases, manifested through the presence of Majorana modes at the boundaries of the open system \cite{Udupa_2021}. Here, we include a non-Hermitian term to extend the topological phase diagram and study its purely non-Hermitian part. For simplicity, we set $\Delta=w$. 
Diagonalizing the Hamiltonian in momentum space results in two bands $E_{\pm}(k)$,
\begin{equation}
    E_{\pm}(k)=\pm\sqrt{w^2+\mu^2-g^2-2w\mu\cos k-2iwg\sin k}, \notag
\end{equation}
for which a line-gap closing occur at $k=0,\pi$. The field theory around those points is then readily obtained from the coefficients in Eq.~\eqref{eq:conH}, $\boldsymbol{\alpha} = (0,\mp w,0)$ and $\boldsymbol{\beta} = (0,-ig,\pm w - \mu)$.
The winding number takes exactly the same form as for the SSH model. It is given by Eq.~\eqref{eq:SSHWinding}, but with $M_{\pm}\equiv (\pm w-\mu)/w$. The inclusion of disorder gives a similar result as obtained for the SSH model.

\paragraph*{Comparison to Numerical Calculations of Disorder in the discrete case.}
We are interested in evaluating the robustness of the energy vorticity when disorder is introduced in the system. For the lattice model, we have \cite{Gong2018,SymTop}
\begin{equation}\label{eq:latticewind}
    \mathcal{W}(E) = -\int_\text{BZ} \frac{\dd k}{2\pi i} \frac{d}{d k} \text{log} \left\{ \text{det} \left[ H(k) - E \right] \right\}.
\end{equation}
Upon taking a continuum limit, it is possible to show that Eq.~\eqref{eq:latticewind} is equal to Eq.~\eqref{eq:effaction} (see SM \cite{SM}). Introducing disorder often leads to a loss of translational invariance. As a result, it is no longer possible to obtain the Bloch Hamiltonian $H(k)$, rendering Eq.~\eqref{eq:latticewind} useless. This problem is solved by introducing a Peierls-like phase in the intercell hopping \cite{Gong2018}, yielding (for the non-Hermitian SSH model)
\begin{align}
    H(\Phi) &= (v-g) \sum_{j=1}^N c^\dagger_{A,j} c_{B,j} + (v+g) \sum_{j=1}^N c^\dagger_{B,j} c_{A,j} \notag \\
    &+ w \sum_{j=1}^N \left( e^{+i(\Phi/N)} c^\dagger_{B,j} c_{A,j+1} + \text{h.c.}\right), \notag
\end{align}
The Hamiltonian $H(\Phi)$ itself is not periodic upon increasing the flux $\Phi$ by $2\pi$, but the quantity $\text{Det}\left[ H(\Phi)\right]$ is \cite{Gong2018}. This allows us to define
\begin{equation}\label{eq:Disvort}
    \mathcal{W}(E) = -\int_0^{2\pi} \frac{\dd \Phi}{2\pi i} \frac{d}{d \Phi} \text{log} \left\{ \text{det} \left[ H(\Phi) - E \right] \right\}.
\end{equation}
While the two expressions for $\mathcal{W}(E)$ look very similar, it is important to realize that the latter is a function of the real-space Hamiltonian. This has the important consequence that we no longer require translational invariance to calculate $\mathcal{W}(E)$. We consider the disorder potential $V = \sum_{j=1}^N V_j\left( c^\dagger_{j,A}c_{j,A}+ c^\dagger_{j,B}c_{j,B}\right)$.
The values of $V_j$ are sampled from a uniform distribution $[-V_0,V_0]$, where $V_0$ is the disorder strength. In Fig.~\ref{fig:disorderwide}(d), we plot the results obtained by averaging over 100 realizations of disorder for the same parameter values used for the continuum model shown in Fig.~\ref{fig:disorderwide}(b). We observe that as the disorder strength increases, the system is driven towards a trivial state, in which the energy vorticity winding number is zero, which agrees with the general features observed from the analytic derivations. Note that this kind of disorder breaks the sublattice  symmetry of the system and leads to an earlier onset of the phase transition. If instead one would implement disorder in the hopping amplitudes in a uniform way, the symmetry would be preserved and the phase transition would occur for higher values of $V_0$. However, as discussed in the SM \cite{SM}, the averaging procedure in the continuum model does not distinguish between these two forms of lattice disorder, limiting the application of the field theory description. Similar results were obtained in the context of Anderson localization using the replica method \cite{AndersonReplica}.

\paragraph*{Conclusions.}
We introduced a generic field-theoretical method to analytically study the effect of disorder in one dimensional two-band non-Hermitian models that feature one or more band closing points. We have shown how the intrinsically non-Hermitian topological phases of these systems are affected by disorder within the SCBA. A non-Hermitian topological invariant naturally arises when coupling the continuum field theory to a background gauge field, and is expressed in terms of a trace over the momentum-space Green's function. The resulting change in this Green's function can then be tracked when we apply averaging procedures in the perturbative expansion. We apply these ideas to the paradigmatic non-Hermitian SSH model, and a model featuring non-Hermitian $s$-wave superconductivity The two models exhibit very similar dispersion relations and are therefore equally influenced by disorder. One would expect these transitions to be extremely sharp, as they are represented by a topological quantity. However, the computations represent averages over many disorder realisations, which smooths out the transitions. This feature is even more prominent in the analytic model. Nonetheless, the results still allow us to capture the fact that a topological phase transition occurs upon the introduction of sufficiently strong disorder, which is expressed by a change in the averaged energy vorticity. 

One might also wonder how disorder affects the skin modes. This has been studied previously using numerical approaches \cite{Gong2018, LonghiDeform, DisorderNHSETransition}, but it would be interesting to investigate whether the current framework provides analytic tools to understand this effect. In addition, one could use the replica method to obtain further insight on the effects of disorder, as done in a study of disordered topological semimetals \cite{DisWeylSM}. The use of the replica method may produce a richer phenomenology and new insights.

Finally, it would be interesting to investigate the effect of interactions in these non-Hermitian topological models. We can apply the same techniques to study their effects on the winding number by simply replacing the bare Green's function with an interacting one. We are confident that the methodology developed here will stimulate further research in this direction. 

We would like to thank R. Arouca and T.H. Hansson for fruitful discussions about the use of field theory for non-Hermitian systems. This publication is part of the project TOPCORE with
project number OCENW.GROOT.2019.048 which is financed by
the Dutch Research Council (NWO).
\\ \\
Authors Anouar Moustaj \& Lumen Eek contributed equally to this work.

\bibliography{mainnew}
\onecolumngrid

\newpage
\appendix

\setcounter{equation}{0}
\setcounter{figure}{0}
\setcounter{table}{0}
\makeatletter
\renewcommand{\theequation}{S\arabic{equation}}
\renewcommand{\thefigure}{S\arabic{figure}}
\renewcommand{\bibnumfmt}[1]{[S#1]}
\renewcommand{\citenumfont}[1]{S#1}

\begin{widetext}
\section*{Supplemental material for `Field Theoretical Study of disorder in Non-Hermitian Topological Models'}
\begin{center}
    \size{11}{Anouar Moustaj, Lumen Eek, and Cristiane Morais Smith}\\
    \size{10}{\textit{Institute for Theoretical Physics, Utrecht University,}}\\
    \textit{Princetonplein 5, 3584CC Utrecht, The Netherlands}\\
\end{center}
\subsection*{Derivation of the energy vorticity}\label{A}
In this section, we will derive the general expression for the energy vorticity, starting from the vacuum-to-vacuum transition amplitude, as introduced in the main text:
\begin{equation}
    Z_{E}[A] = \text{Det} \left[ -i G^{-1}_{0,E} - i(\boldsymbol{\alpha}\cdot\boldsymbol{\sigma})A\right], \notag
\end{equation}
with $G_{0,E}^{-1}(k) = E - (\boldsymbol{\alpha}\cdot \boldsymbol{\sigma})k - \boldsymbol{\beta}\cdot \boldsymbol{\sigma}$. Here $\text{Det}[...]$ denotes taking a determinant over both coordinate and spinor space, while $\text{det}[...]$ only takes the determinant over spinor space. The same convention holds for taking traces. From the definition of the effective action, we then have
\begin{align}
    S_\text{eff} &= -i \log \left\{ Z_{E}[A]/Z_{E}[0] \right\} \notag \\
    &= -i\log\left\{ Z_{E}[A]\right\} + i\log\left\{ Z_{E}[0]\right\}\notag \\
    &= -i \log \left\{ \text{Det} \left[ - i G^{-1}_{0,E} - i(\boldsymbol{\alpha}\cdot\boldsymbol{\sigma})A \right]\right\} + i  \log \left\{ \text{Det} \left[ - i G^{-1}_{0,E} \right]\right\} \notag \\
     &= -i \Tr \left\{ \log \left[ - i G^{-1}_{0,E} - i(\boldsymbol{\alpha}\cdot\boldsymbol{\sigma})A \right]\right\} + i  \Tr \left\{ \log \left[ - i G^{-1}_{0,E} \right]\right\} , \notag
\end{align}
where in the last line we invoked the identity $\Tr\left\{ \log \left[\dots\right] \right\} = \log \left\{ \text{Det} \left[ \dots \right]\right\}$. The first term in the effective action can be expanded up to linear order in $A$
\begin{align}
    S_\text{eff} &= -i \Tr \left\{ \log \left[ - i G^{-1}_{0,E} \left( \mathbb{I} + G_{0,E}(\boldsymbol{\alpha}\cdot\boldsymbol{\sigma})A \right)  \right]\right\} + i  \Tr \left\{ \log \left[ - i G^{-1}_{0,E} \right]\right\} \notag \\
    &= -i \Tr \left\{ \log \left[  \mathbb{I} + G_{0,E}(\boldsymbol{\alpha}\cdot\boldsymbol{\sigma})A  \right]\right\} \notag \\
    &= -i \Tr \left[ G_{0,E}(\boldsymbol{\alpha}\cdot\boldsymbol{\sigma})A \right] + \mathcal{O}(A^2). \notag
\end{align}
In real space, we can write 
\begin{equation}
    G_{0,E}(x,y) = \int \frac{\dd k}{2\pi} G_{0,E}(k) e^{ik(x-y)}, \notag
\end{equation}
allowing us to perform the partial trace over coordinate space
\begin{align}
    S_\text{eff} &= -i\int \dd x \dd y \tr \left[ G_{0,E}(x,y)(\boldsymbol{\alpha}\cdot\boldsymbol{\sigma}) \right] A(y) \delta(x-y) \notag \\
    &= \int \frac{\dd k}{2\pi i} \tr \left[ G_{0,E}(k) (\boldsymbol{\alpha}\cdot\boldsymbol{\sigma}) \right]\int \dd x A(x) \label{eq:appwinding}\\
    &\equiv \mathcal{W}(E) \int \dd x A(x). \notag
\end{align}
Since $G_{0,E}(k)$ is diagonal in momentum space and is written in a basis of Pauli matrices, it is easily found to be
\begin{align}
    G_{0,E}(k) &= \left[E-(\boldsymbol{\alpha}\cdot\boldsymbol{\sigma})k - \boldsymbol{\beta}\cdot\boldsymbol{\sigma} \right]^{-1} \notag \\
    &= \frac{E+(\boldsymbol{\alpha}\cdot\boldsymbol{\sigma})k + \boldsymbol{\beta}\cdot\boldsymbol{\sigma}}{\text{det}\left[E-(\boldsymbol{\alpha}\cdot\boldsymbol{\sigma})k - \boldsymbol{\beta}\cdot\boldsymbol{\sigma} \right]}.\label{eq:appG}
\end{align}
Combining Eq.~\eqref{eq:appwinding} and Eq.~\eqref{eq:appG} together with $\tr[\sigma_i] = 0$ and $\tr[\sigma_i \sigma_j] = 2\delta_{ij}$ then yields
\begin{align}
    \mathcal{W}(E)& = 2\int \frac{\dd k}{2\pi i} \frac{(\boldsymbol{\alpha}\cdot \boldsymbol{\alpha}) k + \boldsymbol{\alpha} \cdot \boldsymbol{\beta}}{\text{det} \left[E - (\boldsymbol{\alpha}\cdot \boldsymbol{\sigma})k - \boldsymbol{\beta}\cdot \boldsymbol{\sigma}\right]}  .\notag \\
    &= 2\int \frac{\dd k}{2\pi i} \frac{||\boldsymbol{\alpha}||^2 k + \boldsymbol{\alpha} \cdot \boldsymbol{\beta}}{E^2 - ||\boldsymbol{\alpha}k + \boldsymbol{\beta}||^2}, \label{eq:genW}
\end{align}
where $||\boldsymbol{\alpha}||$ denotes the complex valued vector norm of $\boldsymbol{\alpha}$.
\subsection*{Derivation of the disorder averaged Green function}\label{B}
In this section, we will show the derivation leading to the disorder averaged Green function. For notational convenience, all sub-scripted $E$'s will be dropped in this section. Let us start from the definition of the Green function
\begin{equation}
    G(x,x') = -\frac{1}{Z} \int \mathcal{D}\psi^\dagger\mathcal{D}\psi \hspace{2pt}\psi(x) \psi^\dagger(x') e^{iS}, \label{eq:defG}
\end{equation}
where the action $S$ is given by 
\begin{equation}
    S= \int \dd x \psi^\dagger (x) \left[E-H_0-\delta V\right] \psi(x).\notag
\end{equation}
Here, $H_0$ is some one-particle quadratic Hamiltonian and $\delta V$ is the disorder potential, which is zero on average. For weak disorder strength, this action can be treated perturbatively,
\begin{align}
    Z &= \int\mathcal{D}\psi^\dagger\mathcal{D}\psi e^{iS} \label{eq:denom} \\
    &= \int\mathcal{D}\psi^\dagger\mathcal{D}\psi \left(1+i\int\dd y \psi^\dagger(y) \delta V(y) \psi(y) + \frac{i^2}{2} \int \dd y \dd z \psi^\dagger(y) \delta V(y) \psi(y) \psi^\dagger(z) \delta V(z) \psi(z) + \dots \right)e^{iS_0}\notag\\
    &= Z_0 \left[ 1 + i\int \dd y G_0(y,y) \delta V(y) + \frac{i^2}{2} \int \dd y \dd z \Big( G_0(y,y) G_0(z,z) -  G_0(z,y) G_0(y,z) \Big) \delta V(y) \delta V(z) \right], \notag
\end{align}
and 
\begin{align}
    -\int\mathcal{D}\psi^\dagger\mathcal{D}&\psi\hspace{2pt} \psi(x)\psi^\dagger(x') e^{iS} \label{eq:num} \\
    &= -\int\mathcal{D}\psi^\dagger\mathcal{D}\psi \hspace{2pt} \psi(x)\psi^\dagger(x') \left(1+i\int\dd y \psi^\dagger(y) \delta V(y) \psi(y)\right.\\
    &+ \left. \frac{i^2}{2} \int \dd y \dd z \psi^\dagger(y)\delta V(y) \psi(y) \psi^\dagger(z) \delta V(z) \psi(z) + \dots \right)e^{iS_0}\notag\\
    &= Z_0\left[G_0(x,x') + i\int\dd y \Big(G_0(x,x')G_0(y,y) - G_0(x,y) G_0(y,x'\Big)\delta V(y)\right.\notag \\
    &+ \left. \frac{i^2}{2} \int \dd y \dd z \Big( G_0(x,x') G_0(y,y) G_0(z,z) + G_0(x,x') G_0(y,z)G_0(z,y) - 2G_0(x,y) G_0(x',y) G_0(z,z) \right. \notag \\
    &+ \left. 2G_0(x,z) G_0(y,x') G_0(z,y) \Big) \delta V(y) \delta V(z) \right] \notag
\end{align}
where $S_0$, $G_0$ and $Z_0$ denote the unperturbed action, Green function and vacuum-to-vacuum transition amplitude, respectively. In the above, Wick's theorem has been used extensively. Substituting Eq.~\eqref{eq:denom} and Eq.~\eqref{eq:num} in Eq.~\eqref{eq:defG} then yields, to second order, in $V(x)$
\begin{align}
    G(x,x') = G_0(x,x') - i \int \dd y G_0(x,y) \delta V(y) G_0(y,x') + i^2 \int \dd y \dd z G_0(x,y) \delta V(y) G_0(y,z) \delta V(z) G_0(z,x'), \notag
\end{align}
or, up to any order in $V(x)$
\begin{equation}
    G(x,x') = G_0(x,x') + \int \dd y \dd z  G_0(x,y) \Sigma(y,z) G(y,x'), \label{eq:fullG}
\end{equation}
where $\Sigma(y,z) = -i \delta V(y) \delta(y-z)$. This follows from the fact that any disconnected Green's function contribution from the numerator gets cancelled by the denominator contribution, at any order, and we can just extend the derivation to all orders. Recall that $\delta V(x)$ is defined as
\begin{equation}
    \delta V(x) = V(x) - \overline{V(x)}, \notag
\end{equation}
with $V(x) = \sum_{i=1}^{N_{imp}} U(x-x_i)$. The disorder average was then defined in the main text as
\begin{equation}
        \overline{V(x)} = \frac{1}{L^{N_{imp}}} \int \dd x_1 \dd x_2 \cdots \dd x_{N_{imp}} V(x). \notag
\end{equation}
By definition we have $\overline{\delta V(x)} = 0$, such that the terms linear in $\delta V(x)$ in Eq.~\eqref{eq:fullG} vanish upon taking the disorder average. This then yields
\begin{equation}
    \overline{G(x,x')} = G_0(x,x') + i^2 \int \dd y \dd z G_0(x,y) G_0(y,z)\overline{\delta V(y) \delta V(z)} \hspace{2pt} \overline{G(y,x')},\notag
\end{equation}
in which we recognise the self-energy
\begin{equation}
    \Sigma(y,z) = - G_0(y,z)\overline{\delta V(y) \delta V(z)}. \label{eq:disself}
\end{equation}
The average of the product of potential terms simplifies to
\begin{equation}
    \overline{\delta V(y) \delta V(z)} = \overline{V(y) V(z)} - \overline{V(y)} \hspace{2pt} \overline{V(z)}. \notag
\end{equation}
Taking delta-peaked disorder, i.e. $U(x-x_i) = U_0 \delta (x-x_i)$, we find $\overline{V(y)} = U_0 n_i$ and
\begin{align}
    \overline{V(y)V(z)} &= \frac{1}{L^{N_{imp}}} \int \dd x_1 \dd x_2 \cdots \dd x_{N_{imp}} U_0^2 \sum_{i=1}^{N_{imp}}\sum_{j=1}^{N_{imp}} \delta(y-x_i)\delta(z-x_j) \notag\\
    &= \frac{U_0^2}{L^{N_{imp}}} \left( L^{N_{imp}-1} N_{imp} \delta(y-z) + L^{N_{imp}-2} N_{imp}(N_{imp}-1)  \right) \notag \\
    &\approx U_0^2 n_i^2 + U_0^2 n_i \delta(y-z), \label{eq:dVdV}
\end{align}
for $N_{imp}\gg1$. This leads to
\begin{equation}
    \Sigma(y,z) = -U_0^2 n_i G_0(y,z) \delta(y-z),\notag
\end{equation}
with constant Fourier components, given by
\begin{align}
    \Sigma(k) = -U_0^2 n_i \int \frac{\dd q }{2\pi} G_0(q). \notag 
\end{align}

\subsection*{Calculations for the non-Hermitian SSH model}\label{C}
\begin{figure}[H]
    \centering
    \includegraphics[width=0.5\textwidth]{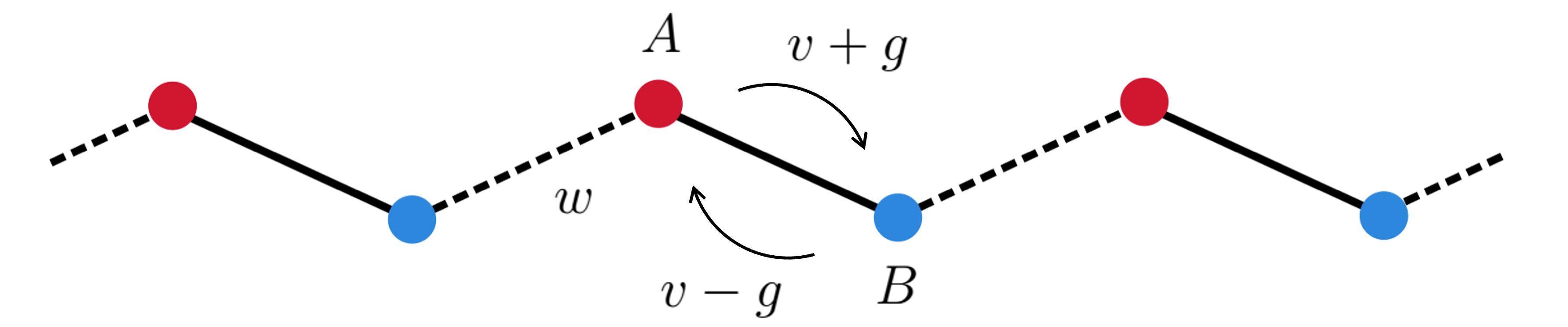}
    \caption{Sketch of the (non-Hermitian) SSH model. Two sites connected by a solid line form a cell. Intercell and intracell hopping occur with  amplitudes $w$ and $v$, respectively. Non-Hermiticity is introduced using the parameter $g$.}
    \label{fig:SSHchain}
\end{figure}
\noindent Fig.~\ref{fig:SSHchain} shows a sketch of the non-Hermitian SSH chain considered in the main text. The SSH Hamiltonian reads 
\begin{equation}
    \hat{H} = \sum_k \mathbf{c}^\dagger_k \begin{pmatrix}
    0 & v-g+w e^{-ik} \\ v + g + w e^{ik} & 0
    \end{pmatrix}  \mathbf{c}_k. \notag
\end{equation}
Expanding the exponential up to linear order allows one to write the matrix as $(\boldsymbol{\alpha}\cdot\boldsymbol{\sigma})k + \boldsymbol{\beta}\cdot\boldsymbol{\sigma}$, with the coefficients
\begin{equation}
     \boldsymbol{\alpha} =  \begin{pmatrix}
     0 \\ \pm w \\ 0 
     \end{pmatrix}, \qquad
     \boldsymbol{\beta} =  \begin{pmatrix}
     v \pm w \\ -ig \\ 0 
     \end{pmatrix}. \notag
\end{equation}
Substituting these coefficients in Eq.\eqref{eq:genW} yields
\begin{align}
    \mathcal{W}_\pm(E) &= 2 \int \frac{\dd k}{2 \pi i} \frac{w^2 k \mp i w g}{E^2 + g^2 - w^2 k^2 \pm 2i w g k - (v\pm w)^2} \notag \\
    &= 2 \int \frac{\dd k}{2 \pi i} \frac{ k \mp i \gamma}{\mathcal{E}^2 + \gamma^2 -  k^2 \pm 2i \gamma k - M_\pm^2} \notag,
\end{align}
where we introduced the scaled parameters $\gamma = g/w$, $\mathcal{E} = E/w$ and $M_\pm = (v\pm w)/w$. Upon rewriting the denominator this turns into
\begin{align}
    \mathcal{W}_\pm(E) &= -2 \int \frac{\dd k}{2 \pi i} \frac{ k \mp i \gamma}{\left[ k \mp i \left( \gamma + \sqrt{M_\pm^2 - \mathcal{E}^2} \right) \right]\left[ k \mp i \left( \gamma - \sqrt{M_\pm^2 - \mathcal{E}^2} \right) \right]} \notag \\
    &= -\int \frac{\dd k}{2 \pi i} \frac{ k \mp i \left( \gamma + \sqrt{M_\pm^2 - \mathcal{E}^2} \right)  +  k \mp i \left( \gamma - \sqrt{M_\pm^2 - \mathcal{E}^2} \right) }{\left[ k \mp i \left( \gamma + \sqrt{M_\pm^2 - \mathcal{E}^2} \right) \right]\left[ k \mp i \left( \gamma - \sqrt{M_\pm^2 - \mathcal{E}^2} \right) \right]} \notag \\
    &= -\int \frac{\dd k}{2 \pi i} \left[ \frac{1}{k \mp i \left( \gamma - \sqrt{M_\pm^2 - \mathcal{E}^2} \right)} + \frac{1}{k \mp i \left( \gamma + \sqrt{M_\pm^2 - \mathcal{E}^2} \right)}\right] \notag \\
    &=-\int \frac{\dd k}{2 \pi i} \left[ \frac{\left(k\mp\Im\sqrt{M^2_\pm-\mathcal{E}^2}\right) \pm i \left( \gamma - \Re\sqrt{M_\pm^2 - \mathcal{E}^2} \right)}{\left(k\mp\Im\sqrt{M^2_\pm-\mathcal{E}^2}\right)^2 + \left( \gamma - \Re\sqrt{M_\pm^2 - \mathcal{E}^2} \right)^2} + \frac{\left(k\pm\Im\sqrt{M^2_\pm-\mathcal{E}^2}\right) \pm i \left( \gamma + \Re\sqrt{M_\pm^2 - \mathcal{E}^2} \right)}{\left(k\pm\Im\sqrt{M^2_\pm-\mathcal{E}^2}\right)^2 + \left( \gamma + \Re\sqrt{M_\pm^2 - \mathcal{E}^2} \right)^2} \right]\notag \\
    &= \mp \frac{1}{2} \left[ \text{sgn} \left( \gamma - \Re \sqrt{M_\pm^2-\mathcal{E}^2}\right) + \text{sgn} \left( \gamma + \Re \sqrt{M_\pm^2-\mathcal{E}^2}\right) \right], \label{eq:suppwind}
\end{align}
where in the last line we used the integral
\begin{equation}
  \int \frac{\dd k}{2 \pi i} \frac{k + i \alpha}{k^2 + \alpha^2} = \frac{1}{2} \text{sgn}\left( \Re \alpha \right). \notag
\end{equation}
In the remainder of this section, we will set $E = 0$, which we argued to be the correct choice in the main text. The phase diagram corresponding to Eq.~\eqref{eq:suppwind} is shown in Fig.~1. It is noteworthy that this invariant renders all phases that are adiabatically connected to the Hermitian model ($g=0$) indistinguishable from each other. This is a result of the purely non-Hermitian nature of the energy vorticity $\mathcal{W}_\pm(E)$.
In order to find the self-energy for delta-peaked disorder, we first evaluate
\begin{align}
    \int \frac{\dd q}{2 \pi} G_0(q) &= -\frac{1}{w} \int \frac{\dd q}{2 \pi} \frac{1}{q^2 + M_\pm^2 - \gamma^2 \mp 2i\gamma q} \begin{pmatrix}
    0 & \mp iq + M_\pm -\gamma \\
    \pm iq + M_\pm + \gamma & 0
    \end{pmatrix} \notag \\
    &= -\frac{1}{w} \int \frac{\dd q}{2 \pi} \frac{\mp i}{\left[(q\mp i(\gamma - M_\pm)\right]\left[(q\mp i(\gamma+M_{\pm})\right]} \begin{pmatrix}
    0 & q \mp i(\gamma- M_\pm) \\
    -q \pm i(\gamma+M_\pm) & 0
    \end{pmatrix} \notag \\
    &= \frac{1}{w} \int \frac{\dd q}{2 \pi i}
    \begin{pmatrix}
    0 & \mp \frac{1}{q \mp i (\gamma + M_\pm)} \\
     \pm \frac{1}{q \mp i (\gamma - M_\pm)} & 0
    \end{pmatrix} \notag \\
    &= \frac{1}{2w} 
    \begin{pmatrix}
    0 & -\text{sgn}\left(\gamma + M_\pm \right)\\
    \text{sgn}\left(\gamma - M_\pm \right) & 0 
    \end{pmatrix}.
\end{align}
Substituting this in the expression for the disorder averaged energy vorticity, we obtain
\begin{align}
    \overline{\mathcal{W}_\pm(0)}&=\int\frac{\dd k}{2\pi i}\tr\left[(1-\Sigma_E(k))^{-1} G_{0,0}(k)(\boldsymbol{\alpha}\cdot\boldsymbol{\sigma})\right] \notag\\
    &= \int \frac{\dd k}{2 \pi i} \frac{1}{1+(U_0^2 n_i/2w)^2 \text{sgn}(\gamma - M_\pm)\text{sgn}(\gamma + M_\pm)} \notag \\
    &\cross \tr \left[ 
    \begin{pmatrix}
    1 & \frac{U_0^2 n_i}{2w}\text{sgn}(\gamma + M_\pm)\\
    - \frac{U_0^2 n_i}{2w}\text{sgn}(\gamma - M_\pm) & 1 
    \end{pmatrix}
    G_{0,0}(k) \left(\boldsymbol{\alpha}\cdot \boldsymbol{\sigma} \right)\right] \notag\\
    &= \frac{1}{1+\left(\frac{U_0^2 n_i}{2w}\right)^2 \text{sgn}(\gamma - M_\pm)\text{sgn}(\gamma + M_\pm)}\int \frac{\dd k}{2 \pi i}\tr \left[ G_{0,0}(k) \left(\boldsymbol{\alpha}\cdot \boldsymbol{\sigma} \right)\right] \notag \\
    &= \frac{\mathcal{W}_\pm(0)}{1+\left(\frac{U_0^2 n_i}{2w}\right)^2 \text{sgn}(\gamma - M_\pm)\text{sgn}(\gamma + M_\pm)},
\end{align}
where in line 2 to 3 we used that the Green function is diagonal in the sublattice sector, such that the off-diagonal contributions to the self-energy vanish upon taking the trace.

\subsection*{Gap structure}
For the non-Hermitian SSH model, the phase transition changes the gap topology from a line-gap to a point-gap and vice versa. This is illustrated in Fig.~\ref{fig:SSHgap}.
\begin{figure}[H]
    \centering
    \includegraphics[width=0.8\textwidth]{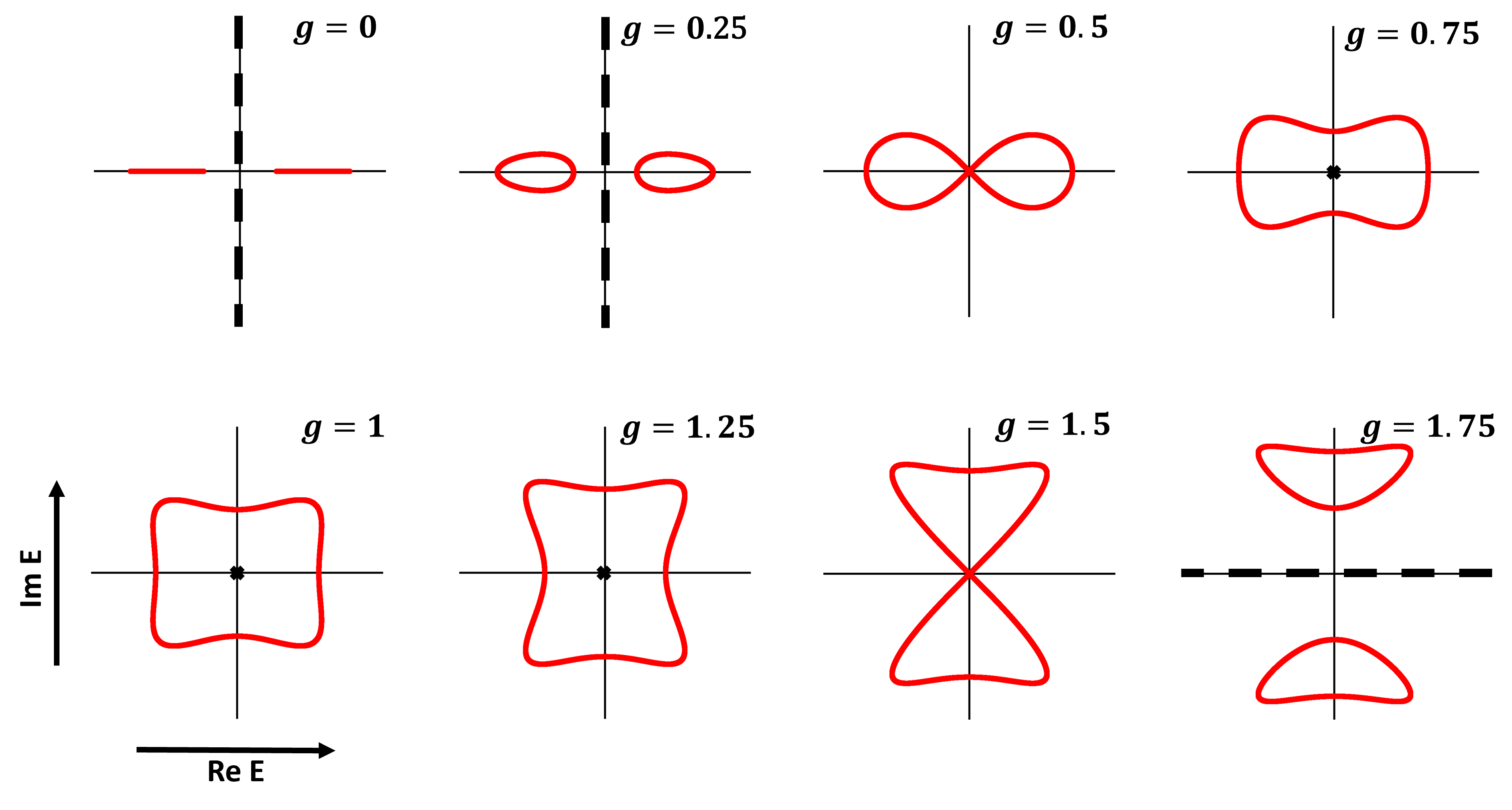}
    \caption{Gap structure of the non-Hermitian SSH model with parameter choices $w=1$ and $v=1/2$. For these values, the gap structure changes at $g=0.5$ and $g=1.5$.}
    \label{fig:SSHgap}
\end{figure}

\subsection*{Energy vorticity in the continuum limit}
Starting from Eq.~(13) of the main text,
\begin{equation*}
    \mathcal{W}(E) = -\int_\text{BZ} \frac{\dd k}{2\pi i} \frac{d}{d k} \text{log} \left\{ \text{det} \left[ H(k) - E \right] \right\},
\end{equation*}
and taking the continuum limit stretches the integration bounds to encompass the whole real line. Furthermore, we should substitute $H(k) = (\boldsymbol{\alpha}\cdot \boldsymbol{\sigma})k + \boldsymbol{\beta}\cdot \boldsymbol{\sigma}$, 
\begin{align*}
    \mathcal{W}(E) &= -\int_{-\infty}^\infty \frac{\dd k}{2\pi i} \frac{d}{d k} \text{log} \left\{ \text{det} \left[ (\boldsymbol{\alpha}\cdot \boldsymbol{\sigma})k + \boldsymbol{\beta}\cdot \boldsymbol{\sigma} - E \right] \right\} \\
    &= -\int_{-\infty}^\infty \frac{\dd k}{2\pi i} \frac{d}{d k} \text{tr} \left\{ \text{log} \left[ (\boldsymbol{\alpha}\cdot \boldsymbol{\sigma})k + \boldsymbol{\beta}\cdot \boldsymbol{\sigma} - E \right] \right\}\\
    &= -\int_{-\infty}^\infty \frac{\dd k}{2\pi i} \text{tr} \left\{ \frac{d}{d k}\text{log} \left[ (\boldsymbol{\alpha}\cdot \boldsymbol{\sigma})k + \boldsymbol{\beta}\cdot \boldsymbol{\sigma} - E \right] \right\}.
\end{align*}
Performing the derivative yields
\begin{equation*}
    \mathcal{W}(E) = -\int_{-\infty}^\infty \frac{\dd k}{2\pi i} \text{tr} \left\{ \left[ (\boldsymbol{\alpha}\cdot \boldsymbol{\sigma})k + \boldsymbol{\beta}\cdot \boldsymbol{\sigma} - E \right]^{-1} (\boldsymbol{\alpha}\cdot \boldsymbol{\sigma}) \right\}.
\end{equation*}
Using $G^{-1}_{0,E}(k) = E - (\boldsymbol{\alpha}\cdot \boldsymbol{\sigma})k - \boldsymbol{\beta}\cdot \boldsymbol{\sigma}$, we can write
\begin{equation*}
    \mathcal{W}(E) = \int_{-\infty}^\infty \frac{\dd k}{2\pi i} \text{tr} \left[ G_{0,E}(k) (\boldsymbol{\alpha}\cdot \boldsymbol{\sigma}) \right],
\end{equation*}
which is equal to Eq.~(3) of the main text.

\subsection*{Off-diagonal disorder implementation}
We will now show that the implementation of disorder in the hopping amplitudes, which results in an off-diagonal disorder potential in the continuum limit,  leads to the same contribution to the  average energy vorticity obtained when considering on-site disorder. To this end, we add a random variable to the amplitudes $v$ and $w$. In the continuum model, this amounts to adding a disorder potential that couples to the fermion field components $\psi_A(x)$ and $\psi_B(x)$, namely:
\begin{equation*}
    H_d=\int dx \boldsymbol{\Psi}^\dagger \delta V(x)\sigma_1\boldsymbol{\Psi}.
\end{equation*}
Upon taking the disorder average, the terms with odd powers of disorder fluctuations vanish and we are left with the nearest-order contribution of the self-energy given by 
\begin{equation}
    \Sigma_E(x,y)=-G_{0,E}(x,y)\overline{\delta V(x)\sigma_1\delta V(y)\sigma_1}. \label{eq:SigE}
\end{equation}
From Eq.~\eqref{eq:SigE}, we see that since $\sigma_1^2=\mathbbm{1}$, we end up with the same contribution as Eq.~\eqref{eq:selfenergy} in the main text. This means that the continuum model cannot distinguish between a sublattice-symmetry preserving hopping disorder and the symmetry-breaking on-site disorder of the lattice model. This sets up a boundary in the current approach.

\subsection*{Self-consistent Born approximation}
In the main text, we showed that the implementation of disorder leads to a different expression for the energy vorticity by correcting the Green function through the self-energy. This self-energy was written in the Born approximation as $\Sigma_0=-G_0\overline{\delta V\delta V}$. If one considers instead the full Green's function, one obtains the SCBA. We can solve it in an iterative way through 
\begin{equation*}
    \Sigma_E^{(n)}(k) = -U_0^2 n_i \int \frac{\dd q}{2\pi} \overline{G^{(n-1)}_E(q)},
\end{equation*}
\begin{equation*}
    \overline{G_E^{(n)}(k)} = \left[ 1 - \Sigma^{(n)}_E(k) \right]^{-1} \overline{G^{(n-1)}_{E}(k)},
\end{equation*}
with $\overline{G_E^{(0)}(k)} = \overline{G_E(k)}$ and $\Sigma_E^{(0)}(k)$ corresponding to the Born approximation. From this, we calculate the SCBA winding number to order $n$:
\begin{equation*}
    \overline{\mathcal{W}^{(n)}_\pm(0)}=\int\frac{dq}{2\pi i}\text{tr}\left\{\left[\prod_{j=1}^n\left(1-\Sigma_0^{(n-j-1)}\right)^{-1}\right]G_0^{(0)}(q)(\boldsymbol{\alpha}\cdot\boldsymbol{\sigma})\right\}.
\end{equation*}
It is found that the general expression for the $n^{\text{th}}$ order winding number is given by that of the zeroth order one, modulated by a rational function of $U_0^2$,
\begin{equation}\label{genwindingcorrections}
    \overline{\mathcal{W}^{(n)}_\pm(0)}=\frac{\mathlarger{\sum}_{j=0}^na^{(n)}_j\left(\frac{U_0^2n_i}{2w}\text{sgn}(\gamma+M_\pm)\text{sgn}(\gamma-M_\pm)\right)^{j}}{\mathlarger{\sum}_{j=0}^{n+2}b^{(n)}_j\left(\frac{U_0^2n_i}{2w}\text{sgn}(\gamma+M_\pm)\text{sgn}(\gamma-M_\pm)\right)^{j}}\mathcal{W}_\pm(0),
\end{equation}
where the coefficients $a^{(n)}_j$ and $b^{(n)}_j$ are renormalized at each iteration. One can see in Fig.~\ref{fig:consistent} that the transition becomes sharper as the number of iterations increases. We have also added the analytic expressions of the first four corrections in table \ref{tab:corrections}.
\begin{figure}[H]
    \centering
    \includegraphics[width=0.5\textwidth]{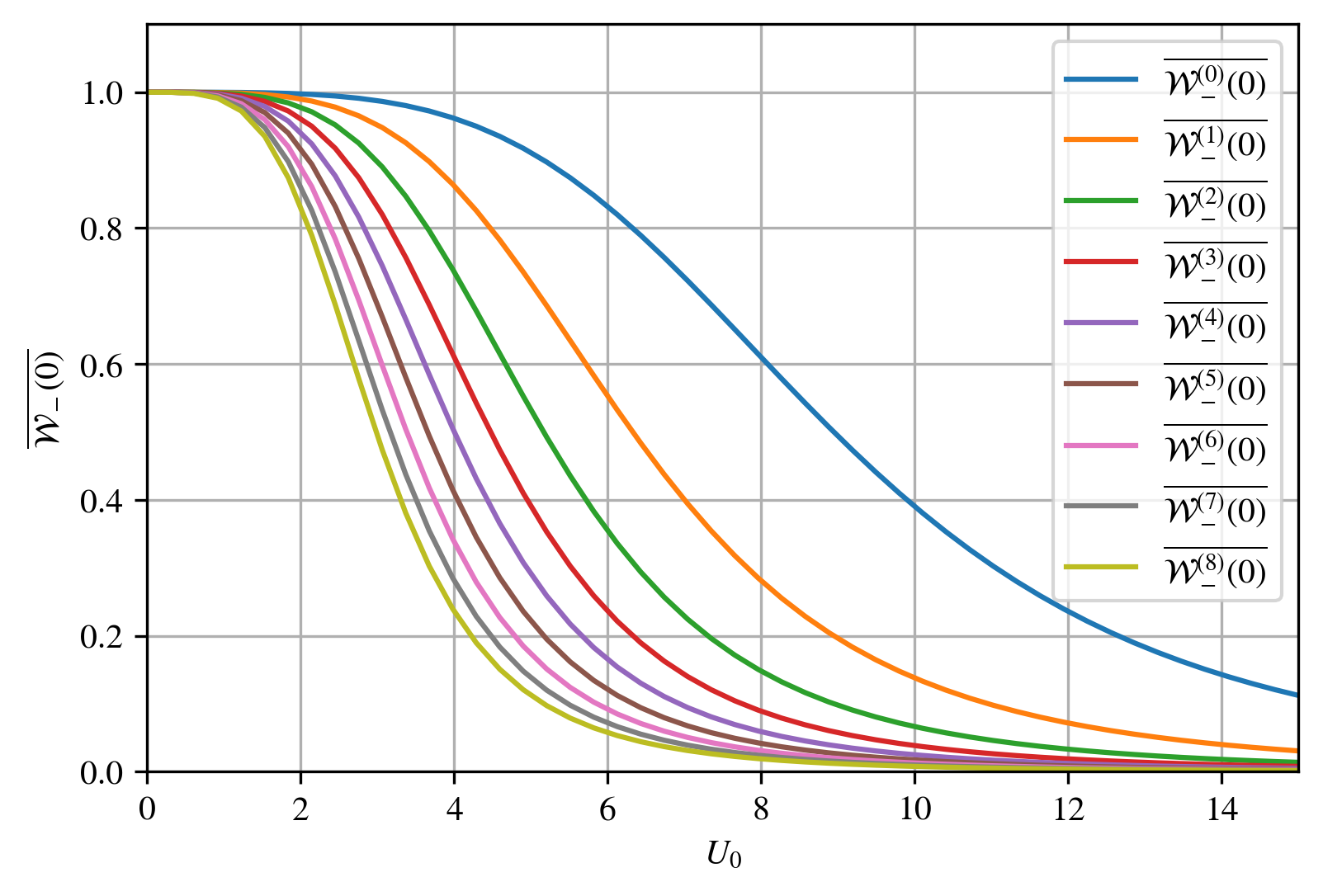}
    \caption{Averaged winding number $\overline{\mathcal{W}_-(0)}$ evaluated for the first eight iterations of the SCBA, for parameters $(v,w,g,n_i) = (0.5,1,1,0.025)$.}
    \label{fig:consistent}
\end{figure}
\begin{table}[!hbt]
    \centering
    \begin{tabular}{|c|c|c|}
        \hline
          Order & Value & Additional expressions \\
          \hline\hline
          $\displaystyle\mathcal{W}^{(0)}_\pm$  & $\displaystyle\mp \frac{1}{2} \left[ \text{sgn} \left( \gamma - M_\pm\right) + \text{sgn} \left( \gamma + M_\pm\right) \right]$ & \\ \hline
          $\displaystyle\mathcal{W}^{(1)}_\pm$ & $\displaystyle\frac{1}{1+\mathcal{U}^2 S}\mathcal{W}^{(0)}_\pm$ & \makecell{$\displaystyle S\equiv\text{sgn}(\gamma - M_\pm)\text{sgn}(\gamma + M_\pm)$ \\ $\displaystyle\mathcal{U}\equiv\left(\frac{U_0^2 n_i}{2w}\right)$} \\ \hline
          $\displaystyle\mathcal{W}^{(2)}_\pm$ & $\displaystyle\frac{1+\mathcal{U}^2 S}{\left(1+2\mathcal{U}^2S\right)^2+\mathcal{U}^2S}\mathcal{W}^{(0)}_\pm$ & \\ \hline 
          $\displaystyle\mathcal{W}^{(3)}_\pm$ & $\displaystyle\frac{\xi}{\left(\xi+2\mathcal{U}^2 S\right)^2+\mathcal{U}^2S}\mathcal{W}^{(0)}_\pm$, & $\displaystyle\xi=\frac{\left(1+2\mathcal{U}^2 S\right)^2+\mathcal{U}^2S}{1+\mathcal{U}^2S}$ \\ \hline 
          $\displaystyle\mathcal{W}^{(4)}_\pm$ & $\displaystyle\frac{\xi\Omega}{\left(\Omega+\varphi\mathcal{U}^2S\right)^2+\xi^2\mathcal{U}^2S}\mathcal{W}^{(0)}_\pm $  & \makecell{$\displaystyle\Omega\equiv \left(\xi+2\mathcal{U}^2S\right)^2+\mathcal{U}^2S$  \\ $\displaystyle\varphi\equiv 2\xi+4\mathcal{U}^2S+1$ } \\ \hline 
    \end{tabular}
    \caption{First four corrections to the winding number, as given by the SCBA. Notice that the expressions follow the general formula given by Eq.\eqref{genwindingcorrections}. }
    \label{tab:corrections}
\end{table}
In the main text, we set $n_i=0.025$ to compare the discrete and continuum models in a consistent way. Fig.~\ref{fig:consistentni} shows the averaged energy vorticity in the SCBA as a function of impurity density. One observes that higher impurity density leads to an earlier onset of the phase transition. This behaviour is to be expected, because a higher impurity density yields a more disordered system. 
\begin{figure}[H]
    \centering
    \includegraphics[width=0.5\textwidth]{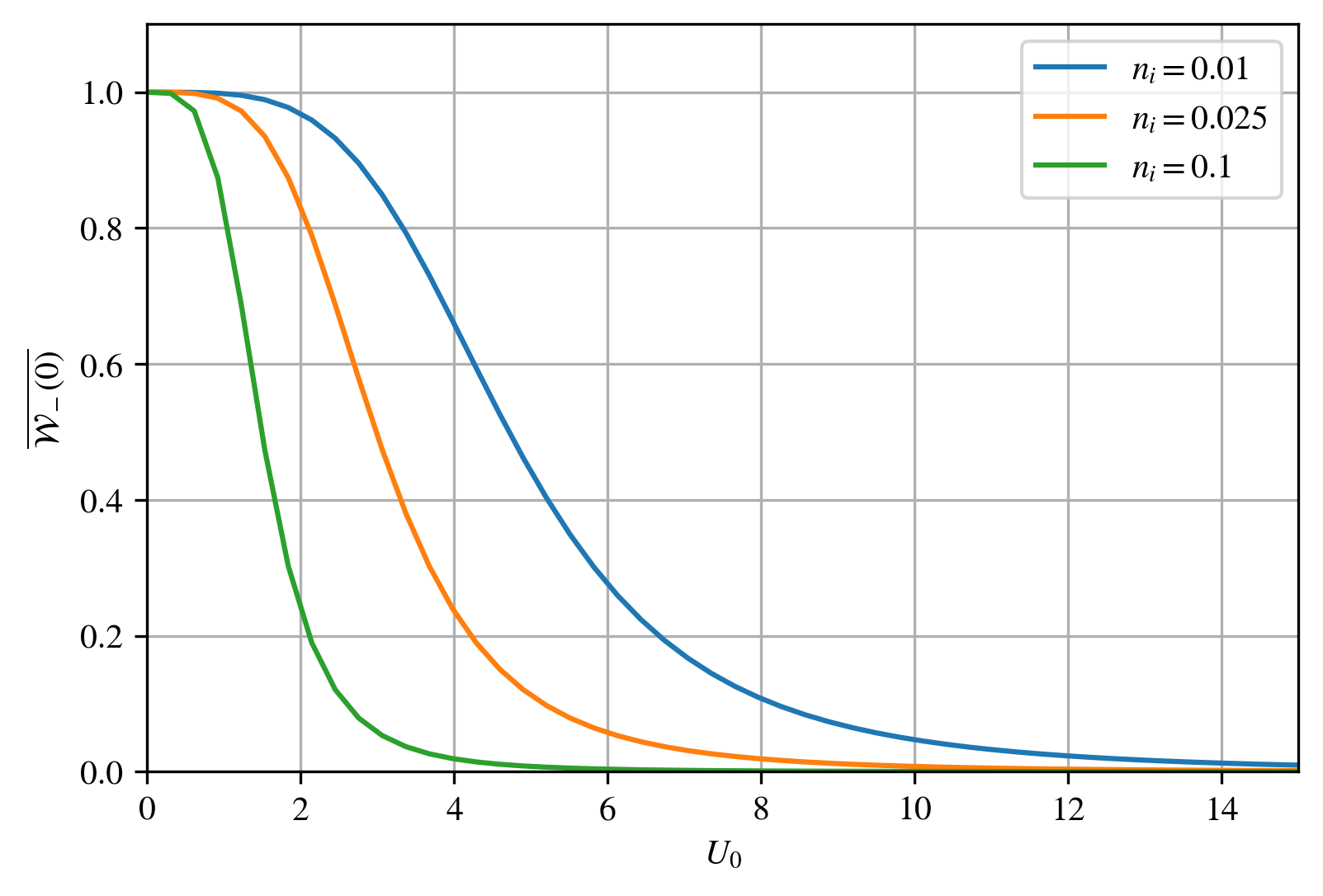}
    \caption{Dependence of $\overline{\mathcal{W}_-(0)}$ on the impurity density $n_i$  in the SCBA for parameters $(v,w,g) = (0.5,1,1)$.}

    \label{fig:consistentni}
\end{figure}

\end{widetext}

\end{document}